\newcommand{\Cro}{{\rm Cro}}
\newcommand{\cro}{{\rm cro}}
\newcommand{\cI}{{\sl cI}}

\newcommand{\CI}{{\rm CI}}

\newcommand{\lacZ}{{\sl lacZ}}

\newcommand{\recA}{{\sl recA}}
\newcommand{\recAminus}{{\sl recA$-$}}

\newcommand{\RecA}{{\rm RecA}}
\newcommand{\Ecoli}{{\sl E.~coli}}
\newcommand{\OL      }{$O_{\rm L}$}
\newcommand{\OR      }{$O_{\rm R}$}

\newcommand{\ORone   }{$O_{\rm R}1$}

\newcommand{\ORtwo   }{$O_{\rm R}2$}

\newcommand{\ORthree }{$O_{\rm R}3$}

\newcommand{\PR      }{$P_{\rm R}$}

\newcommand{\PRM     }{$P_{\rm RM}$}

\newcommand{\PRE     }{$P_{\rm RE}$}

\newcommand{\RR      }{$R_{\rm R}$}
\newcommand{\RRmath      }{R_{\rm R}}
\newcommand{\RRMU     }{$R_{\rm RM}^u$}
\newcommand{\RRM     }{$R_{\rm RM}$}
\newcommand{\RRMmath     }{R_{\rm RM}}
\newcommand{\RRMUmath     }{R_{\rm RM}^u}

\newcommand{\SRM    }{$S_{\cI}$}
\newcommand{\SRMmath    }{S_{\cI}}
\newcommand{\Scro    }{$S_{\cro}$}
\newcommand{\Scromath    }{S_{\cro}}

\documentstyle[12pt]{article}

\begin{document}

\title{Stability Puzzles in Phage $\lambda$}
\author{Erik Aurell$^1$, Stanley Brown$^2$, 
Johan Johanson$^3$ and Kim Sneppen$^4$}
\maketitle
\begin{center}
\begin{tabular}{ll}
$^1$ & Dept. of Mathematics, Stockholm University,
SE-106 91 Stockholm, Sweden \\
$^2$ & Dept. of Molecular Cell Biology, University of Copenhagen, \\
&$\quad$ \O ster Farimagsgade 2A, DK-1353 Copenhagen K, Denmark \\
$^3$ & Dept. of Microelectronics and Nanoscience, CTH, SE-412 96 
G\"oteborg, Sweden \\
$^4$ & NORDITA, Blegdamsvej 17, DK-2100 Copenhagen, Denmark
\end{tabular}
\end{center}
\vspace{0.5cm}
\noindent
\begin{tabular}{ll}
Keywords: &  Gene regulation, stability, robustness, \\
		&	phage lambda, lysogeny \\ 
		&			\\
Corresponding author & K. Sneppen, NORDITA \\
		     & Blegdamsvej~17,  DK-2100 Copenhagen, Denmark \\
		     & phone~+45~3532~5365, fax~+45~3538~9157 \\
		     & e-mail~sneppen@nbi.dk \\
		     &			
\end{tabular} 

\newpage

\begin{abstract}
In the absence of RecA-mediated cleavage of the repressor, 
the $\lambda$ prophage is exceptionally stable. 
In fact, the repressed state is then more stable than the gene 
encoding the repressor.  We develop a mathematical treatment 
that predicts the stability of such epigenetic states from 
affinities of the molecular components. 
We apply the model to the behavior of recently published 
mutants of \OR\ and find that their observed stability indicates 
that the current view of the \OR\ switch is incomplete.
The approach described here should be 
generally applicable to the stability of expressed states.  
\end{abstract}

\section{Introduction}
\label{s:introduction}
The lysogeny maintenance switch in 
phage $\lambda$ can be viewed as one of the simplest examples
on the molecular level of computation, 
command and control in a living system.  
If, following infection of the bacterium {\sl Escherichia coli},
the virus enters the lysogenic pathway,
it represses its developmental functions,
in response to a small set of sensory inputs, and
integrates its DNA into the host chromosome.
In this state the prophage 
may be passively replicated for many generations of \Ecoli.
\\\\
Established lysogeny is maintained
by the protein \CI\ which blocks operators \OR\ and \OL\
on the $\lambda$ DNA,
and thereby prevents transcription of all lytic genes including {\sl cro} 
(Ptashne, 1992).
In lysogeny the \CI\ concentration
functions as a sensor
of the state of the bacterium: if DNA is damaged
the protease activity of \RecA\ is activated,
leading to degradation of \CI. A low \CI\ concentration
allows for transcription of the lytic genes, starting
with {\sl cro}, the product of which is the protein \Cro.
\\\\
According to the picture so beautifully conveyed by Ptashne 
(Ptashne, 1992),
the $\lambda$ lysogeny maintenance switch is centered around operator \OR,
and consists of three binding sites 
\ORone, \ORtwo\ and \ORthree, each of which can be occupied by
either a \Cro\ dimer or a \CI\ dimer. 
As illustrated in Fig.~1 
these three binding sites control the activity of
two promoters \PRM\ and \PR\ for respectively \cI\ and {\sl cro} 
transcription.  Transcription of {\sl cro} starts at
\PR, which partly overlaps \ORone\ and \ORtwo.
Transcription of \cI\ starts at \PRM, which partly overlaps
\ORtwo\ and \ORthree.
The affinity of RNA polymerase for the two promoters,
and subsequent production of the two proteins, depends on
the pattern of \Cro\ and \CI\ bound to the three operator
sites and thereby establishes lysogeny 
with about 200 \CI\ molecules per bacterium.
If, however, \CI\ concentration
becomes sufficiently low, the increased production of \Cro\ throws
the switch to lysis.
\\\\
There does not seem to be a quantitative theoretical understanding
of how stable the lysogenic state is to spontaneous switching.
We will demonstrate that the large stability observed
experimentally puts constraints on the mechanism of the switch,
constraints which we explore in a quantitative model.
We will show that in order to span the timescales
from single molecule on--off binding times 
of order about 10 sec, to the stability of order
$10^{12}$ sec, a simple model, where lysis  
occurs every time \cro\ is transcribed, is inadequate.
We elaborate a dynamical model that takes
into account all known mechanisms of \OR,
including in particular nonspecific binding between
\Cro\ dimers and \Ecoli\ DNA.
We can then reproduce the observed high stability. 
However, we will see that the recently reported
robustness of the switch against
mutational changes of the operators 
(Little {\it et al.}, 1999)
suggest that an additional mechanism of
stability must be present.
\\\\
Bistable switches are expected to be important functional
units in many systems of control of gene expression 
(Monot \& Jacob, 1961, Glass \& Kauffman, 1973). 
One general consequence of the 
mathematical model presented here is, that in living
systems such switches can often only be of finite
stability, on account of the small number of molecules
involved. The model is hence also relevant to the
stability of data storage in synthetic gene-regulatory networks
with similar characteristics, in particular if intended for use in
biotechnology or gene therapy (Gardner et al, 2000).

\section{Experiments on lysogen stability}
\label{s:experimental-measurement-of-stability}

Repression of bacteriophage $\lambda$
can be eliminated either by cleavage of the repressor 
by the RecA protease 
(reviewed by Roberts \& Devoret, 1983)
or by the absence of repressor at the operators.  
The frequency of spontaneous 
induction in strains deleted for the recA gene has recently been 
reported.  Rozanov {\it et al.}, (1998) 
found the frequency of infectious 
centers to be $1.1 \cdot 10^{-7}$.  
Little {\it et al.}, (1999) 
measured the ratio of released phage per lysogen and the burst size.  
They estimate the frequency of one cell to switch to the lytic state to be 
$4 \cdot 10^{-7}$ per generation, assuming the switch occurred one hour before lysis.  
Little et al. verified the lysogens contained only a single prophage by PCR.  
Rozanov {\it et al.}, (1998) 
do not report 
testing the number of prophages in the lysogens they used.  
We repeated these measurements using 
single lysogens (Powell {\it et al.}, 1994) 
of MG1655 (Jensen, 1993)
containing the 
del(srl-recA)306::Tn10 allele 
(Csonka {\it et al.} 1979).
We measured the frequency of switching to the lytic state by measuring 
the appearance of turbid plaques in a limiting dilution experiment
at 37$^{o}$ in YT broth 
(Miller, 1972).
We find that
lysed cells to appear with frequency between 1.0 and 
$1.8 \cdot 10^{-8}$, which, 
if we include the same correction factor as 
(Little {\it et al.}, 1999)
to account for the 
increase in cell numbers during the period of phage development, implies
a rate in the range of $3.5 \cdot 10^{-8}$ 
and $6.0 \cdot 10^{-8}$ per generation and cell.  
Thus, all three sets of
measurements fall within a factor of $10$, in spite of the use of different
strain backgrounds.
\\\\
Recent results of J. Little (Univ. of Arizona, personal communication) 
show that at 37$^o$C about 99\% of the
turbid plaques are not wild type, but instead a marginally stable
PRM240 mutant. As a consequence, Little estimates that wild type
$\lambda$ phages inside \Ecoli\ cells grown on nutrient rich medium
have a frequency of spontaneous lysis 
that does not exceed $2 \cdot 10^{-9}$ per generation and cell.

\section{Elements of a model}
\label{s:stochastic-chemistry-of-the-switch}
The considerations in this section are in part closely
similar to the model of (Shea \& Ackers, 1985). 
The main difference is that our interest is in longer
time scales, and in the stability of the switch against
spontaneous stochastic fluctuations. 
We will therefore begin with
a discussion of characteristic time scales, 
then discuss
chemistry, production and decay. 
The stochastic element is included
in a definite model, see next section.
\\\\
{\bf Time scales:}
The measured half lives of \CI\ and \Cro\ dissociation from \ORone\ 
at $22^o$C 
are respectively $34$ sec for \CI\ 
(Nelson \& Sauer, 1985),
and varies from $1200$ sec for \Cro\ 
(Jana {\it et al.}, 1998) 
dissociation from \OR\ alone 
down to 20 sec 
(Kim {\it et al.} (1987))
from OR flanked by 1kb of $\lambda-$DNA.
At equilibrium dissociation balances association, and thus
the time it takes a \CI\ or \Cro\
dimer to bind to the operator is
\begin{equation}
\tau_{1} \;=\; \tau_{off} \; V \; [1M] \;  \exp(\Delta G/RT)
\label{eq:tau-off-rate}
\end{equation}
where $\Delta G$ is the free energies of binding,
which for \CI\ and \Cro\ are tabulated in Table~\ref{table:Ackers-Gs},
and $RT=0.617\, \hbox{kcal/mol}$. 
Assuming a bacterial volume of
$V= 2 \cdot 10^{-15}\,l$ 
(Bremmer {\it et. al.}, 1996)
we have $V [1M] =12 \cdot 10^{8}$, and 
can then estimate the association rates of single molecules
to be of order tens of seconds, for both \CI\ and \Cro.
This is compatible with
diffusion-limited association into a region of size
$L=1\, \hbox{nm}$, at a rate determined by
the Smoluchowski equation (Berg {\it et al.}, 1982)
$\tau_{1} \;=\; V/(4 \pi D L)$, using a 
diffusion constant $D=10^{-11} m^2/s$, similar to
that of other proteins in bacterial cytoplasma 
(Elowitz {\it et al.}, 1999).
The association rate from $N$ molecules is proportional to $N$,
and thus for a bacterium with 10-200 \CI\ and \Cro\ molecules
association events will occur at a frequency of 
fractions of a second,
for both \CI\ and \Cro.
\\\\
The shortest time scale in which we are actually interested
is that which it takes for a switching event to take place.
This cannot be much less than those of significant
changes in \CI\ and \Cro\ concentrations,
which are of order
one bacterial generation, $34$ minutes in the strains
used in (Little {\it et al.}, 1999).
Hence, all binding-unbinding events of \Cro\ and \CI\
can be assumed to
take place in homeostatic equilibrium.
In addition, we need to characterize RNA
polymerase binding and
initiation of transcription from
promotor sites \PRM\ and \PR. 
As discussed below (see {\bf Production})
these events also have to occur on time scales
much shorter than one bacterial generation,
and can therefore be described by their overall
rates. We note that a more detailed description
of RNA polymerase action may be necessary
to resolve the 
shorter timescales associated to the lysis-lysogeny entry decision,
as discussed in modelling by McAdams, Arkin and Ross (1997,1998).
\vspace{0.5cm}

\noindent
{\bf Chemistry:}
We assume that  
\CI\ and \Cro\ molecules in the cell are in 
homeostatic equilibrium.
This does not mean that there is always the same
number of \CI\ dimers bound to the operators at any particular time.
These numbers are fluctuating, and the equilibrium assumption gives
the size of these fluctuations.
The key inputs 
are \CI\ and \Cro\ dimerization constants and 
the Gibbs free energies for their bindings
to the three operator sites \ORone, \ORtwo\ and \ORthree\
(Koblan \& Ackers, (1992); Takeda {\it et al.}, 1989,
Kim {\it et al.} (1987), Jana {\it et al.} (1997,1998)),
see Table~\ref{table:Ackers-Gs}.
\\\\
We remark that these free energies are taken from in vitro studies,
and that the in vivo conditions could be different,
e.g. the measured protein-DNA affinities could depend 
sensitively on the ions present in the buffer solutions.
On the other hand, in vivo the effects of such changes should
be compensated for, as e.g. changed $KCl$
concentrations are by 
putrescine (Capp {\it et al.} (1996)) and other ions and
crowding effects
(Record {\it et al.} (1998))
The data quoted in Table~\ref{table:Ackers-Gs} was obtained
at $KCl$ concentration of 200mM, 
which most closely resembles in vivo conditions.
Parameter sensitivity is analysed separately below.
\\\\
Following Ackers {\it et al.} (1982), 
we encode a state $s$ of 
\CI\ and/or \Cro\ bound to \OR\ by three numbers
$(i,j,k)$ referring to respectively
\ORthree\, \ORtwo\ and \ORone. The coding is $0$
if the corresponding site is free, $1$ if the site is occupied
by a \CI\ dimer, and $2$ if the site is occupied by a \Cro\ dimer.
The
probability of a state $s$
with $i_s$ \CI\ dimers and $j_s$ \Cro\ dimers
bound to \OR\ is in the 
grand canonical approach of Shea and Ackers
(Shea \& Ackers, 1985)
\begin{equation}
P_R(s) = {\cal N}^{-1}
[\CI\ ]^{i_s} [\Cro\ ]^{j_s}\; e^{-G(s)/T}
\label{eq:grand-canonical-formula}
\end{equation}
The normalization constant ${\cal N}^{-1}$ is
determined by summing over $s$.
The free dimer concentrations
($[\CI\ ]$ and $[\Cro\ ]$) are equal to 
the numbers of free dimers ($n_D$) per volume, 
where the free dimer numbers are given by
\begin{eqnarray}
2 n_D &=& N - n_M - 2 {\cal N}_{chr} n_U \nonumber \\
&& - 2 \cdot {\cal N}_{\lambda} 
\left( \sum_s i_s P_R(s) + \sum_{s'} i_{s'} P_L(s') \right)
\label{eq:dimer-number}
\end{eqnarray}
where $N$ is the total number of molecules of this type,
${\cal N}_{\lambda}$  and ${\cal N}_{chr}$ the average
number of copies present of the $\lambda$ genome and of the full
\Ecoli\ chromosome, 
$n_U$ the number of dimers bound nonspecifically
to one chromosome $DNA$ and $n_M$ the number of free monomers.
We assume free dimers to be in equilibrium with
free monomers and with dimers bound nonspecifically to
DNA, such that $n_M$ and  $n_U$ can be expressed
in terms of $n_D$, volume, and dimerization and association
constants.
In this paper we take ${\cal N}_{\lambda}$  and ${\cal N}_{chr}$
both equal to $3$~\cite{Bremmer}.
In our analysis we have mostly assumed that \CI\
does not bind nonspecifically to DNA, while
\Cro\ does so with considerable affinity.
The effect of a putative nonspecific \CI\ binding to
DNA is analysed separately,
see Table 1 and 
\ref{s:appendix}.
Note that we also include occupancy of \OL\
in order to account for finite depletion of \CI, which has 
some significance for the fractional binding at \OR\ when
the total number of \CI\ molecules is small.  
Parameters for \OL\ are listed in caption to Table 1.
\\\\
In Table~\ref{table:Ackers-Gs} we have calculated 
${\cal P}(s)=$ ${\cal P}(s\, | \, N_{\CI}=200,N_{\Cro}=0)$.
Of particular interest for the stability are the states 
in which \cro\ can be transcribed, i.e. $(000)$, $(100)$ and $(200)$.
The numbers in the table show that 
\cro\ is always transcribed at some rate
in lysogeny, which if every \cro\ transcription induced 
lysis, would destabilize the lysogenic state within a few generations.
In Fig.~2a we show $P(0,0,0)+P(1,0,0)+P(2,0,0)$ as function of $N_{\CI}$ 
at $N_{Cro}=0$. 
We remark that Fig.~2a 
corresponds to the scaling of \PR\ promoter strength
in the in vitro experiment of Hawley \& McClure (1983). 
The small gap in \CI\ number before full activation of \PRM\ 
at low \CI\ in Fig.~2a originates in part from
finite system filling of \OR\ and \OL.
\\\\
\noindent
{\bf Production:}
\CI\ and \Cro\ are produced from mRNA transcripts
of \cI\ and \cro, which are initiated from promotor
sites  \PRM\ and \PR.
The rate of initiation of transcription
from \PRM\ when stimulated by \CI\ bound
to \ORtwo\ is denoted \RRM, and when not
stimulated \RRMU. The
number of \CI\ molecules produced per transcript 
is \SRM, and overall expected rate of production
of \CI\ is
\begin{eqnarray}
\hbox{Overall \CI\ rate} &=& \RRMmath \SRMmath 
	\left( P(010)+P(011)+P(012) \right) +
	\nonumber \\
&& \RRMUmath \SRMmath ( P(000)+P(001)+P(002)+ \nonumber \\
&& P(020)+P(021)+P(022) 
	)
\label{eq:CI-production-rate}
\end{eqnarray} 
For the relative values of the coefficients, 
Hawley \& McClure (1982) report \RRMU\  to be
equal to $\RRMmath/11$.
The absolute values are not known, and are
likely to at least be proportional to the concentration
of RNA polymerase in the cell. We discuss how we determine
\RRM\ below.
According to
(Shean \& Gottesman, 1992)
the number of \CI\ molecules produced from one transcript of \cI\
is a factor $20 - 70$
smaller compared to \lacZ.
Thus \SRM\ is small, between one and five, and we take
\SRM\ to be one.
\\\\
As there are about 200
\CI\ in a lysogenic cell, and as each transcription
only results in a small number of \CI\ molecules,
transcription has to be initiated from \PRM\
many times per generation in lysogeny.
We can therefore safely assume
that the time scales for RNA polymerase to
find the operator sites is also much less than one
generation. This justifies our description 
of production by the overall rates, without
distinguishing RNA polymerase binding,
the rate of formation of open complexes, and
possible temporary stalling of RNA polymerase
at promotor sites.
Fig.~2b shows average \CI\  production according to
(\ref{eq:CI-production-rate}) as function of
\CI\ number in the cell, volume held constant.
We note that the overall behaviour resembles the in vitro
measurements of Hawley \& McClure (1983).
\\\\
The rate of initiation of transcription
from promoter \PR\ is denoted \RR\ 
and the number of \Cro\ molecules per transcript \Scro. 
From Ringquist {\it et al.}, 1992
one can estimate \Scro\ to be 51\% of ideal lacZ
and (Kennel \& Riezman, 1977)
thus allow a lower estimate
of \Scro\ to be $20$.
Thus, in contrast to
the many small production events of \CI\ ,
\Cro\ production is intermittent. 
The ratio of \RR\ to \RRM\  has been reported to
be between $1.3$ and $20$, and depends strongly
on RNA polymerase concentration in the cell
(Hawley \& McClure, 1982).
In practice we determine \RRM\ by balancing \CI\
production and decay, as discussed below,
and use \RR\ as a free parameter.
We end this part by stressing 
that \RR, \RRM, and \RRMU\ parametrize the total production
rates from all copies of $\lambda$ DNA in the cell.
The real rates of transcription 
per \OR\ complex will generally be smaller, e.g. 
about a factor ${\cal N}_{\lambda}\;\sim \; 3$ 
smaller at high growth rates.
\\\\
\noindent
{\bf Decay:}
Concentrations decay due to dilution and degradation.
For \CI\ we only take into account dilution from cell growth
and division, while for 
\Cro\ we include in addition an 
{\it in vivo} half life time $t_{cro}$ of about 
one hour at 37$^{o}$ (Pakula {\it et al.}, 1986),
which is significant compared to {\it e.g.} generation time of $34$ minutes
in the experiments of Little {\it et al.}, 1999.
\\\\
\noindent
{\bf The steady state:}
Balance between production and decay of \CI\ over
one generation, where production is given by
(\ref{eq:CI-production-rate}) and decay by
$\frac{N_{\CI}}{t_{life}}$,
gives one equation for the parameter \RRM, if
the total numbers of \CI\ and \Cro\ in lysogeny
are assumed known. 
Fig.~2b displays the steady state as the intersection
between the linear dilution and the \PRM\ activity curve,
with 200 \CI\ and no \Cro\ molecules in the system.
\\\\
The average number of \Cro\ molecules
in lysogens has not been measured,
but in the models considered here this number is not zero.
Equation (\ref{eq:Cro-production-rate}) below
gives the average rate of \Cro\ production.
If we assume that \RR\ is not very different from \RRM,
as Shea \& Ackers (1985),
table~\ref{table:Ackers-Gs} fixes the rate of \cro\
transcription in lysogeny to be once in 
five to ten generations, and if we assume \RR\ to
be an order of magnitude larger than \RRM, then 
\cro\ is transcribed on average once per generation.
In both cases, this production is balanced by
\Cro\ decay to give an average number of \Cro.
Hence, the balance of only \CI\ production
and decay does not completely
determine \RRM, since the probabilities depend parametrically
on the \Cro\ concentrations, which in turn depend on \RR.
However, the implied dependence of \RRM\
on \RR\ is not very pronounced. 
In the model presented below we fit 
\RRM\ and \RR\ to data, where the value
of $0.115$ for \RRM\
is not very sensitive to changes in model assumptions.

\section{The model}
\label{a:Kims-model}
The basic time scale in our model is one
bacterial generation, where we resolve shorter
times around  \cro\ transcription events
when necessary.
A simulation goes as follows.
Each bacterial generation starts with
$N_{\CI}$ total number of \CI, and $N_{\Cro}$ total
number of \Cro.
From these one computes by equation (\ref{eq:grand-canonical-formula})
the occupation probabilities $P_s$ of the states $s$ of \CI\
and \Cro\ dimers bound to \OR, and from these
the average \CI\ and \Cro\ production rates
$f_{\CI}$ and $f_{\Cro}$. The first is computed by
equation (\ref{eq:CI-production-rate}) above,
and the second by
\begin{eqnarray}
\hbox{Overall \Cro\ rate} &=& 
\Scromath \RRmath \left(P_{000}+P_{100}+P_{200}\right)
\label{eq:Cro-production-rate}
\end{eqnarray} 
\noindent
\CI\ production is assumed continuous, since 
\SRM\ is about one.
\Cro\ production is on the other hand treated as discrete
events.
In each generation the bacterial volume $V$ is assumed to start at
$V_{start}=\frac{2}{3}V_{ave}$, and then to grow linearly 
until it has doubled. $V_{ave}= 2 \cdot 10^{-15} l$ is the average volume over
one generation. The growth of bacterial DNA is treated similarly,
with an average of 3 DNA chromosomes ($\sim 15 \cdot 10^6$ base pairs per cell). 
\vspace{0.5cm}

\noindent

\noindent
{\bf 1)} 
	The rates $f_{\CI}$ and $f_{\Cro}$ are computed
	with the current values of $N_{\CI}$, $N_{\Cro}$
	and $V$.
	\cro\ transcription events are assumed Poisson
	distributed with mean waiting time $\Scromath/f_{\Cro}$.
	A random time $t_r$ is drawn from this distribution.
	If this time is less than the remaining time of
	the current bacterial generation,
	one \cro\ transcription will occur, and we continue
	the simulation under point 2b) below. 
	Else, no more \cro\ transcription will take place in
	this generation, and we continue under 2a).	
	
\vspace{0.5cm}
\noindent
{\bf 2a)}
	The time since beginning of the generation, or since
	last \cro\ transcription in the generation, is $t$. 
	We update \CI\ numbers by the expected number
	produced in the interval $t$, $f_{\CI}\cdot t$,
	plus $\xi$, a random
	number drawn from a Gaussian distribution with zero
	mean and variance $f_{\CI}\cdot t$.
	We update \Cro\ number by 
	removing each \Cro\ molecule with a probability
	$p=p(t)= 2^{-t/t_{\Cro}}$, and then proceed to 
	cell division under point 3) below.

\vspace{0.5cm}
\noindent
{\bf 2b)} 
	We update \CI\ number and degrade \Cro\ number
	as above, 2a), but under time $t_r$.
	The number of \Cro\ molecules produced from this
	\cro\ transcription is a random variable
	drawn from a Poisson distribution of mean \Scro,
	and is added to the remaining \Cro.
	We now return back to point 1) above for the
	remaining time of the current generation.

\vspace{0.5cm}
\noindent
{\bf 3)}
	This is the stage of cell division.
 	The volume of a daughter cell is restarted 
	at $V_{start}$, and 
	will contain each \CI\ and \Cro\ molecule
	of the mother cell with probability one half.
	If \CI\ number has shrunk below a threshold, 
	a switch to the lytic pathway has
	occurred, and we leave the loop.	
	Otherwise we return to 1) with 
	the new values of \CI\ and \Cro.
\vspace{0.5cm}

\noindent
The expected number of generations before lysis is 
calculated by an average over at least 10 independent trajectories.
The value of the threshold in \CI\ number is
unimportant, provided it is chosen sufficiently low.
In the model presented here lysis practically always
follows if $N_{\CI}$ goes below $20$.

\section{Results}
\label{a:Results}
{\bf Standard parameter behaviour:}
The two parameters \RRM\ and \RR\ have been adjusted such that
the average number of \CI\ molecules in lysogeny is $200$,
and the mean rate of spontaneous lysis is
$\sim 2 \cdot 10^{-9}$ generations. The other parameters values
are discussed above in section~\ref{s:stochastic-chemistry-of-the-switch}.
The values thus obtained are
$\RRMmath=0.115$sec$^{-1}$ and $\RRmath=0.30$sec$^{-1}$.
The ratio between these two rates is hence about three,
well within the limits set by
Hawley \& McClure (1982)
but larger than the ratio 1.3 used in the model by
(Shea \& Ackers, 1985).
\\\\
For these parameters we show in Fig.~3 the last
few hundred generations before lysis in a typical simulation.
The numbers of \CI\ and \Cro\ fluctuate for a long time around
the metastable lysogenic state.
A particularly large fluctuation (of both \CI\ and \Cro) then builds
up over a few generations, leading to lysis.
This kind of switching event is similar to 
the escape over an activation barrier as 
in the well-known diffusion model of chemical reactions. 
Thus, spontaneous lysis is the result of a number of unlikely
events happening in a given short time interval. 
If each event has probability
$p<1$, and lysis depends on $n$ events, 
the resulting frequency of lysis
is $\sim p^n$. This frequency is therefore very sensitive to $p$,
that means to parameters in the model.
\\\\
In Fig.~4 we explore the activation barrier picture 
in a more detailed way by computing 
the probability that
the system over time visits various values of
$N_{\CI}$ and $N_{\Cro}$. The distribution has a maximum
around the lysogenic state, as it should.
From  there to the origin, along
the $N_{\CI}$ axis, the probability becomes almost vanishingly
small, which reflects the fact that a fluctuation of \CI\ alone,
without \Cro, cannot induce lysis.
The switch instead happens along a ridge in the
$(N_{\CI},N_{\Cro})$ plane, where probability does not
go down so quickly. In Fig.~4b we examine the tail of the
distribution in Fig.~4a. One observes a saddle for high \Cro\ number
and $N_{CI}$ about $20-25$.
For the parameter values used here, 
lysis follows with 50\% probability,
as soon as $N_{\CI}$ goes below $23$. It is
natural to assume that lysogeny is separated from
lysis by a basin boundary (a separatrix), which runs 
close to parallel to the $N_{\Cro}$ axis
around the saddle point.
The value of \Cro\ at the saddle can be varied from 
about a hundred to well over a thousand in the model,
and depends sensitively on the numerical values of
\Cro\ affinities to specific and nonspecific DNA.
\\\\
{\bf Parameter sensitivity:}
A main objective to build a model for stability is to address which
key quantities that determines the stability. 
As we have already discussed, stability is governed
by near simultaneous occurrences of many rare events.
Consequently, the stability will be very sensitive to the probability
of these events, which in turn depend sensitively on parameters,
in particular of the parameters describing \Cro\ production in the cell.
We now examine parameter sensitivity, i.e. lysis frequency as function
of production rates and binding affinities.
The least uncertain combination is the product $\SRMmath\cdot\RRMmath$,
which is to good accuracy
determined by the number of \CI\
in lysogeny, known to be between 180 and 350 for growth in rich medium. 
The individual terms in this product are less well known,
but only influence the dynamics through the relative amplitude
of noise in \CI\ production, which will be proportional
to the square root of \SRM, all else equal.
This uncertainty is much less than what stems from \Cro,
as we describe next, and will therefore be ignored.
\\\\
In Table~\ref{table:results} the top row refers 
to a reference set-up, and the 
three following rows show the effects
of varying, subsequently, 
\Cro\ degradation rate $t_{\Cro}^{-1}$,
\RR, and finally \RR\ and \Scro\  keeping the product
$\RRmath\cdot \Scromath$ fixed. 
Thus, with given binding affinities
the mean rate of switching increases with any of
\Scro, \RR\ and $t_{\Cro}$, but by
varying two of these parameters in opposite
directions, one may also reproduce the measured stability.
Varying different binding affinities
we find that stability primarily is determined by
the free energy difference between \Cro\ binding to \ORthree\ and \Cro\
binding to nonspecific DNA. In Fig. 5 we show the dependence
of stability with this difference, for three different values of \RR.
At low lysis frequency, 
a four-fold change in \RR\ corresponds to a little less than a 
$RT \ln 4$ change in binding energy difference
(approximately equal to $0.9\hbox{kcal/mol}$) 
Changes in \Cro\ affinities to \ORone\ and \ORtwo\ do not influence
stability significantly. 
\\\\
\CI\ influences the
stability of a lysogen 
by binding to \ORone\ and \ORtwo, which
determines the
fraction of time \PR\ is open
(see Fig.~1.).
Increasing for instance binding to these sites by $0.5\hbox{kcal/mol}$
means that the
probability that \PR\ is open is reduced from 0.000071 to 0.000013, 
a five-fold weaker activation of \PR. 
\CI\ also indirectly influences the  
stability of a lysogen through the binding
to \ORthree, which determines the fraction of time
\PRM\ is open. To balance decay through dilution at given
total number of \CI, such
a change must be compensated by an increase in 
the product  $\RRMmath\dot\SRMmath$.
An increase in this must then be matched by a change 
in \RR\ to maintain the measured stability.
Thus any \CI\ affinity increase can be compensated
by a corresponding increase in \RR.
\\\\
The conclusion so far is that with 
present parameter information
the model is compatible 
with observed rate of spontaneous
lysis in \recAminus\ \Ecoli\ strains, but we have
to look elsewhere for a more stringent test, given the
present lack of knowledge of especially parameters related to \Cro.
\\\\
{\bf The Little mutants:}
The last section of the Table~\ref{table:results} shows the mean
rate of switching in variants of the model
corresponding to the recently reported 
lysogenic state in $\lambda$ mutants, where either
operator \ORone\ is replaced by a copy of \ORthree, or vice versa
(Little {\it et al.}, 1999).
Thus while wild type (321)
has an \OR\ site made up of
\ORthree \ORtwo \ORone, the mutant labelled
mutant 121 has  \ORone \ORtwo \ORone\ at \OR, 
and mutant 323   \ORthree \ORtwo \ORthree.
Varying parameters related to \Cro\ and promoter strengths we
find it impossible to reproduce simultaneously the measured stability
of wild type \recAminus\ lysogens and the existence of stable lysogens in these mutants.
For 323 the cause of decreased stablity is demonstrated in Fig.~2a, which shows 
that \PR\ is activated much more than in wild type.
For 121, on the other hand, the lack of stability is
reflected in a much lower \CI\ level in lysogens,
as demonstrated in Fig 2b. Thus the two mutants demonstrate two
mechanisms
of destabilization, one through enhanced \Cro, the
other through depleted \CI.   
\\\\
When adjusting \RR\ down by a factor 10
we can obtain stable lysogens for 121,
and if by a factor 100, then also for 323.
In both cases the stability of wild-type increases
enormously, and does not match the supposed
lysis frequency of $2 \cdot 10^{-9}$ per cell and
generation.
These parameter changes allow us however to analyse 
what lysogens in the mutants would look like, as
in the last six lines of Table 2.
We see that the number of \CI\ for 121 is 15-20\% of wild-type,
while for 323 the number is 50-60\%.
Both of these numbers are similar
to the \CI\ levels reported by Little {\it et al.} (1999).
These \CI\ levels can also be found 
by balancing production and decay of \CI\ only,
as in Fig.~2b.
In further qualitative agreement with experiment we
observe that in spite of having more \CI, the 121 mutant is markedly
more stable than 323. This is mainly caused by the much larger 
probability for open \PR\ in 323, as seen from Fig.~2a.
\\\\
Finally we report that the above conclusions remain valid also
if \CI\ binds significantly to nonspecific DNA, see appendix:
In that case we may simultaneously fit stability of wild-type 
and 121, but not wild-type and 323.

\section{Discussion}
\label{s:discussion}
The switch to lysis in our model is essentially a first exit time
problem, in a system influenced by a combination of deterministic and
stochastic forces. A well-known analogy is 
thermal escape of a particle from a potential
(Kramers, (1940); H\"anggi {\it et al.}, 1990),
a model of chemical reactions with activation
energies.
There the rate of escape depends exponentially on the height
of the potential barrier, and is therefore very sensitive to
small changes in the potential.
Our case is not exactly in Kramers' form, but the analogy is
nevertheless illuminating. We also find that the rate of spontaneous
switching depends very sensitively on model parameters.
In fact, the negative of the logarithm of the probability from Fig.~4 
can be identified with the Wentzel-Freidlin quasi-potential
(Freidlin \& Wentzell, 1984), (Maier \& Stein, 1997),
which plays the same role in this more gene\-ral exit time problem
as the potential in Kramers' problem.
An analogy to temperature is on the other hand
the noises associated to
typical fluctuations of \CI\ and \Cro\ numbers in a lysogen.
We intend to return to these
questions in a future publication.
\\\\  
The large stability of the lysogenic state of the $\lambda$
puts constraints on working mechanisms of the switch.
First, we have shown that a simple model, where every transcription
of \cro\ leads to lysis, only provides stability for a few
generations. A mechanism must therefore exist which stabilizes
the switch against most transcriptions of \cro. We have explored
a straight-forward model, where lysogens are stabilized
by a spontaneous lysis that require a number of \cro\ translations
within each of 3-5 subsequent bacteria generations.
Thereby the small stability arising from a single \cro\
translation is raised to a high power, implying
a very rare spontaneous breakdown of lysogeny.
The need for multiple \cro\ translations for break down
of lysogen stability is associated to
the following mechanisms: (i) \Cro\ bound to operators is in
homeostatic equilibrium, and \CI\ production is therefore 
possible with \Cro\ present; (ii) \Cro\ is degraded and diluted
over time; (iii) \Cro\ binds also to nonspecific DNA, with
significant affinity. Measured lysogenic stability of 
wild-type \recAminus\ \Ecoli\ can then be reproduced
in the model.
\\\\
The actual stability depends sensitively on the parameters
in the model, and we thus predict a stability which 
changes dramatically with changes in e.g. growth conditions.
The model allows for quantitative examination of mutants,
by changing parameters in the model, and we have 
in this sense investigated the
\OR\ mutants reported by 
Little {\it et al.}, 1999.
Although the model can reproduce the ratios of
\CI\ numbers in mutant and wild-type
lysogens, and of the stabilities of each
one of wild-type, 121 and 323 mutants,
it fails for all parameter values to reproduce
simultaneously all of the stabilities. The discrepancy
is largest between wild type and 323.
This suggests that some additional mechanism outside the \OR\ complex
contributes to the stability.
Origins of such a mechanism could be either
an increased repression of \Cro\ production by \CI\
relative to our model, or, conversely, a
decreased repression of \CI\ production by \Cro.
Examples of such mechanisms are:
i) additional \cI\ transcription directed from \PRE;
ii) an unknown interaction between \Cro\ and \CI\ that
allows \Cro\ to stabilize lysogens;
iii) \CI\ mediated binding between \OL\ and \OR\, if that
would significantly repress \PR;
iv) that \Cro\ is unable to block \PRM\ completely,
even when bound to \ORthree.
\\\\
The first mechanism i) would give a second
role to \PRE\ in repressor maintenance,
while mechanism ii) was suggested already by
(Eisen {\it et al.}, 1982) based on a study 
of the Hyp phenotype. 
An argument for mechanism iii) is the recent report by 
R\'evet et al (1999), which shows a fourfold
increase in repression of \PR\ in a plasmid construction
involving two operator complexes, both of which can bind
\CI\ dimers, and one of which overlaps with \PR\ as
in wild-type $\lambda$ \OR.
These mechanisms are of course not exclusive. 
\\\\
In summary, we have developed a quantitative model
for the stability of lysogens. The model builds a direct
connection between processes, their affinities and the resulting
stability. The model allows for quantitative numerical
tests of genetic feed-back mechanisms on the molecular
level, and points to possible shortcomings
in the standard model of $\lambda$ lysogeny
maintenance switch.
Finally, we wish to point out again that the model developed here
speaks directly to the stability of data storage in
biocomputing (Gardner, {\it et al.} 2000).

\section*{Appendix: Consequences of nonspecific \CI\ binding}
\label{s:appendix}
(Koblan {\it et al}, 1992) report
that binding per base pair between
\CI\ dimers and nonspecific DNA is at least
$9\hbox{kcal/mol}$ weaker than binding
between \CI\ and \ORthree.
This sets a limit on
nonspecific binding to be at most 
$-3.5 \hbox{kcal/(mol$\cdot$ bp)}$.
Here we consider the effects of a
$-3.0\hbox{kcal/(mol$\cdot$ bp)}$ nonspecific binding
between \CI\ dimers and DNA, see also Table 1. 
The $2\cdot 10^{-9}$ stability of lysogens with 
$N_{\CI}=200$ is then obtained with
$\RRMmath=0.085/sec$ and $\RRmath = 0.02/sec$. 
In this state \ORthree\ is 20\% occupied in lysogeny, in agreement with 
(Maurer {\it et al.}, 1980).
When we also apply these parameters to the 
121 and 323 mutants we find that 121 can form stable lysogens 
(with $N_{CI}(121) \approx 40$ and lysis frequency
in the range $10^{-3}$ to $10^{-5}$),
whereas 323 fails to do so.
This lack of stability of 323 in the model fails to
reproduce the stability observed by (Little {\it et al.}, 1999).

\section*{Acknowledgements}
\label{s:acknowledgements}
We thank D.~Court, I.~Dodd, B.~Egan, H.~Eisen, J.~Little,
M.~Mossing and P.~Muratore-Ginanneschi
for discussions and valuable comments.
We thank I. Dodd abd B. Egan for suggesting
repression by \Cro\ may be incomplete. 
We thank the Swedish Natural Research Council for support under grant
M-AA/FU/MA 01778-333 (E.A), and the Danish research council 
for financial support (S.B.).
We also thank K.F. Jensen and B.E. Uhlin for gifts of 
MG1655 and the \recA\ allele, respectively.

\begin{figure}[ht]
\caption{
Right operator complex, \OR, consisting of the three operators
\ORone\ , \ORtwo\  and \ORthree. \cI\ is transcribed when \ORthree\
is free and \ORtwo\ is occupied by \CI.
\cro\ is transcribed when both \ORtwo\ and \ORone\ is free.
\CI\ dimers bind cooperatively to \ORone\ and \ORtwo.
}
\end{figure}

\begin{figure}[ht]
\caption{ 
{\bf a)} Probability of open \cro\ promoter, $P(\cdot 00)=P(000)+P(100)+P(200)$,
	as function of \CI\ number
	in {\sl E.coli} cell of volume $2 \cdot \mu m^3$.
	The dashed lines marked
	121 and 323 show repression
        for the two \OR\ mutants investigated
	 by J. Little (Little {\it et al}, 1999).
{\bf b)} Activity of \PRM\ in units of activity when
        \ORthree\ is free and \ORtwo\ is occupied by \CI.
        Same notation as in a). For relative activity of nonstimulated \PRM\ 
        promoter we use the data of (Hawley \& McClure, 1982).
}
\end{figure}

\begin{figure}[h]
\caption{Time course of last few hundreds of generations
	before lysis in a typical simulation with the
	standard set of parameters.
	The average switching rate is $1.4 \cdot 10^{-9}$
	per cell and generation. 
        The real level of \Cro\ in a lysogenic cell is unknown.
	In our model, almost all of the \Cro\ present is bound
	nonspecifically to DNA in lysogeny, the total level
	could be changed by at least a factor 10 by changing
	the (uncertain) difference in \Cro\ affinity
	to \ORthree\ and to nonspecific DNA.
}
\end{figure}

\begin{figure}[h]
\caption{ {\bf a) }Probability to be at various values of 
	\CI\ and \Cro\ number in the cell during lysogeny.
        Due to finite number of samples, one cannot see the
        rare fluctuations leading to lysis.
        {\bf b)} As in a), but limited to the rare events 
        where $N_{Cro}>50$. 
}
\end{figure}

\begin{figure}[b]
\caption{Frequency of spontaneous
        lysis ($f$) as function of the affinity ratio 
        $r=\exp(\Delta\Delta G/RT)$ between specific and nonspecific
        \Cro\ binding.
        The full line is for standard setup (e.g. $\RRmath =0.30$/s),
        and one observe a large relative change in $f$ 
	with $r$,  implying that the adjustment in bindings
        that leads from high to very high stability is small.
        The +) show effect of increased nonspecific \Cro\ -- DNA binding by 
	$1\hbox{kcal/(mol$\cdot$bp)}$, and its
	similarity to the full line marks that no free \Cro\ are present.
        The dashed lines show lysis frequency with respectively
        $\RRmath =1.2$/s and $\RRmath =0.075$/s. 
}
\end{figure}

\newpage

\begin{table}[ht]
\begin{tabular}{|l|l|l||l|l|}\hline
 & & $\Delta G_{\CI-DNA}=$  &  $\Delta G_{\CI-DNA}=$  \\
                            & & $0.0\,\hbox{kcal/(mol$\cdot$ bp)}$ & 
	$-3.0\,\hbox{kcal/(mol$\cdot$ bp)}$ \\\hline
State $s$ &  $\Delta G$ $[\hbox{kcal}/\hbox{mol}]$. & P(s,N=200) & P(s,N=200) \\\hline
$(000)$    & $   0.0  $ & 0.00005 & 0.00063\\
$(001)$    & $   -12.5$ & 0.00301 & 0.01269\\
$(010)$    & $   -10.5$ & 0.00012 & 0.00050\\
$(100)$    & $    -9.5$ & 0.00002 & 0.00010\\
$(011)$    & $   -25.7$ & 0.59291 & 0.80189\\
$(101)$    & $   -22.0$ & 0.00147 & 0.00199\\
$(110)$    & $   -22.9$ & 0.00632 & 0.00855\\
$(111)$    & $   -35.4$ & 0.39606 & 0.17366\\
\CI\ dimer & $   -11.1$ &        &        \\\hline
$(002)$    & $  -14.4 $ &        &        \\
$(020)$    & $  -13.1 $ &        &       \\
$(200)$    & $  -15.5 $ &        &        \\\hline
\Cro\ dimer& $   -7.0 $ &        &       \\
$\Delta G_{\Cro-DNA}$ & $  -6.5/ $bp&        &     \\\hline
\end{tabular}
\caption{
\CI\ and \Cro\ dimer affinities 
to \ORone, \ORtwo\ and \ORthree\ after,
respectively, (Koblan \& Ackers, 1992),
and
(Takeda {\it et al.}, 1989, 1992),
(Jana {\it et al.} 1997, 1998)
and (Kim {\it et al.} 1987).
A major source of uncertainty of in vivo
\Cro\ bindings is that
all in vitro experiments for \Cro\
have been performed at or below room temperature.
The \CI\ dimerization is from 
(Koblan \& Ackers 1991),
\Cro\ dimerization
from (Jana {\it et al.} 1997),
and nonspecific \Cro\ dimer binding to
DNA from (Kim {\it et al.}, 1987).
The third and fourth columns
display
the probabilities of different states
of binding of \CI\
to the sites in \OR, with different assumed nonspecific
affinity of \CI\ to DNA.
For binding of single \CI\ dimers
to left operators the
relative affinities compared to \ORone\ ($\Delta G_R(0,0,1)$) are
$\Delta\Delta G_L(s) = \{ 1.0,0.8,-0.2 \} $ in units of
$\hbox{kcal/mol}$, where $s$ are the
states $(0,0,1)$,  $(0,1,0)$ and  $(1,0,0)$, respectively.
The relative affinities of binding of single \Cro\ dimers are
similarly $\Delta\Delta G_L(s) = \{ -0.1,0.5,1.0 \} $,
where the reference energy is $\Delta G_R(0,0,2)$
(Takeda {\it et al.}). 
}
\label{table:Ackers-Gs}
\end{table}

\newpage 
\begin{table}[ht]
\begin{tabular}{|l|l|l|l||l|l|l|}\hline
\OR\ &  \RR & \Scro  & $t_{\Cro}$  & \CI\ &  \Cro\ & lysis frequency \\
type & [s$^{-1}$]& number  & [s] & number  & number & [generations]$^{-1}$ \\\hline
321 & 0.30   & 20 & 3600  & 200 & 0.8& $1.4 \cdot 10^{-9}$ \\\hline
321 & 0.30   & 20 & 7200  & 200 & 0.8& $1.7 \cdot 10^{-8}$ \\
321 & 0.60   & 20 & 3600  & 200 & 1.4& $3 \cdot 10^{-7}$ \\
321 & 0.30   & 40 & 3600  & 200 & 1.6& $2 \cdot 10^{-5}$ \\\hline
121 & 0.30   & 20 & 3600  &     &    & $>0.1$ \\
323 & 0.30   & 20 & 3600  &     &    & $>0.1$ \\
121 & 0.030  & 20 & 3600  &  32 & 8  & $1.3 \cdot 10^{-6}$ \\
323 & 0.030  & 20 & 3600  &     &    & $ \sim 0.1$ \\
121$^*$ & 0.030  & 20 & 3600  &  32 & 8  & $4 \cdot 10^{-6}$ \\
323$^*$ & 0.030  & 20 & 3600  &  85 & 34 & $0.04$ \\
121 & 0.0030 & 20 & 3600  &  33 & 1  & $5 \cdot 10^{-7}$ \\
323 & 0.0030 & 20 & 3600  & 122 & 2  & $1.5 \cdot 10^{-6}$ \\\hline
\end{tabular}
\caption{
Switching stabilities
and \CI\ and \Cro\ numbers
as function of model parameters.
for wild type and the Little mutants.
Top row is the standard configuration,
as described in the main text.
Second through fourth row explores the
changes in lysis frequency and average number
of \Cro\ present in wild type
by varying \RR, \Scro\ and  $t_{\Cro}$.
Fifth to twelfth row show the
changes in Little 121 and 323
mutants upon changing \RR.
The rows marked 121* and 323* explore an
additional change of \RRMU, the rate of
unstimulated transcription from \PRM\
to $\RRMUmath = 0.30 \RRMmath$.
The standard parameter value (without *)
is $\RRMUmath = \RRMmath/11$.
We observe that the 323 mutants fail to
stabilize unless \RR\ is decreased a
hundredfold. The stability of 121 is increased
a million times (compare row 5 and 11), as
is also wild-type.
}
\label{table:results}
\end{table}

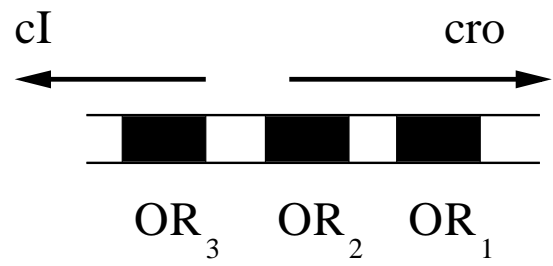
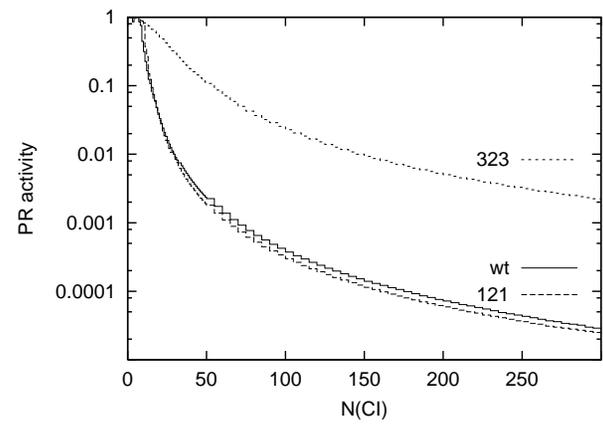
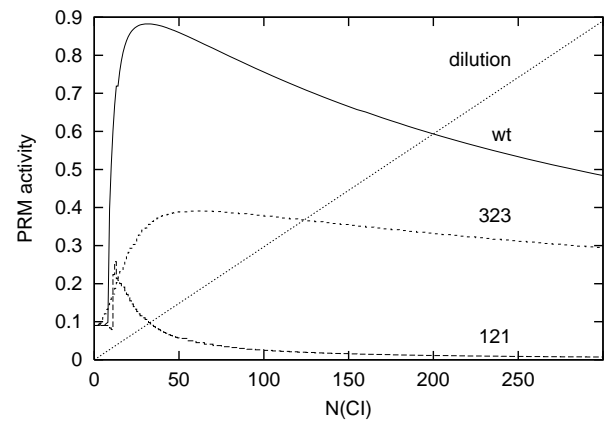
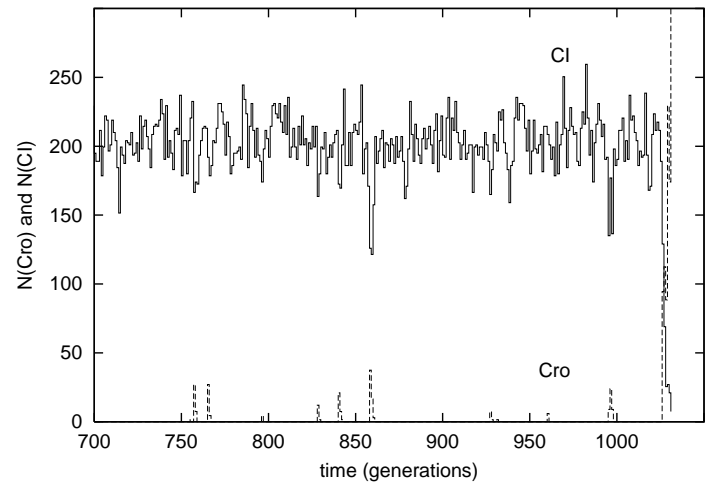
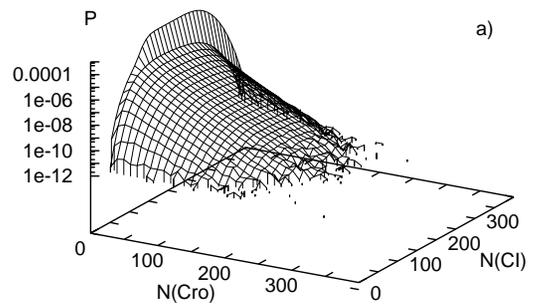
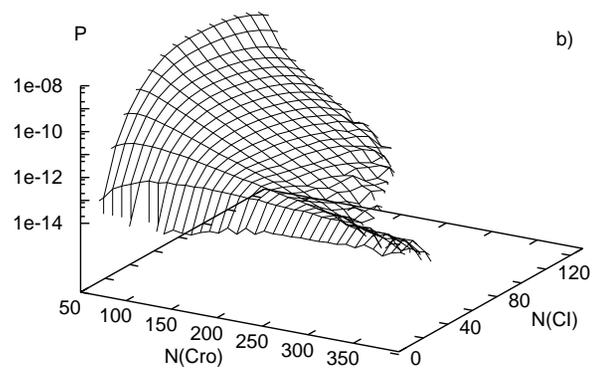
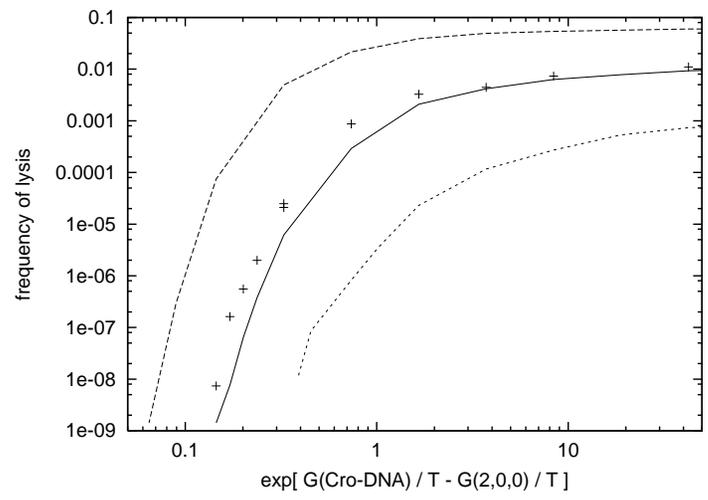
\end{document}